\documentclass[12pt]{article}
\newcommand{\beq}{\begin{equation}}
\newcommand{\ee}{\end{equation}}
\newcommand{\bea}{\begin{eqnarray}}
\newcommand{\eea}{\end{eqnarray}}
\newcommand{\nn}{\nonumber}

       \newcommand{\bv}{{\mathbf v}}

       \newcommand{\bB}{{\mathbf B}}

\usepackage{color}
\usepackage{amsmath}
\usepackage{epsfig}
\usepackage{epstopdf}
\usepackage{xcolor}
\usepackage{babel}
\usepackage[iso-8859-7]{inputenc}
\usepackage{graphicx} 
\usepackage{booktabs} 
\usepackage{blindtext}
\usepackage[a4paper,
            bindingoffset=0.2in,
            left=1in,
            right=1in,
            top=1in,
            bottom=1in,
            footskip=.25in]{geometry}

\usepackage{hyperref}
\hypersetup{
    colorlinks=true,
    linkcolor=red,
    filecolor=magenta,      
    urlcolor=cyan,
    citecolor = cyan,
    }
\input epsf.sty

\begin{document}

\begin{center}
\noindent
{\large \bf A similarity reduction

\vspace{2mm}
of the generalized Grad-Shafranov equation}\vspace{4mm}

A. I. Kuiroukidis$^1$, D. A. Kaltsas$^{1,2}$ and G. N. Throumoulopoulos$^1$ \vspace{4mm}

$^1$Department of Physics, University of Ioannina, GR 451 10 Ioannina, Greece  \vspace{4mm}

$^2$Department of Physics, International Hellenic University, Kavala, Greece, GR 654 04

\vspace{3mm}
Emails: a.kuirouk@uoi.gr,\ kaltsas.d.a@gmail.com, \  gthroum@uoi.gr
\end{center}


\begin{abstract}

We extend previous work [Y. E. Litvinenko, Phys. Plasmas {\bf 17}, 074502 (2010)]
on a direct method for finding similarity reductions of partial differential
equations such as the Grad-Shafranov equation (GSE), to the case of the
generalized Grad-Shafranov equation (GGSE) with arbitrary incompressible flow.  Several families of analytic solutions are constructed, the generalized
Solov\'ev solution being a particular case, which contain both the classical and non-classical group-invariant solutions to the GGSE.  Those solutions can describe a variety of equilibrium configurations pertinent to toroidal magnetically confined plasmas and planetary magnetospheres.

\end{abstract}

\newpage


\section{Introduction}\

The analysis of similarity reductions of partial differential
equations plays a crucial role in many physical applications.
Some time ago in \cite{litvi} a direct method for finding similarity
reductions of the Grad-Shafranov equation (GSE) 
has been introduced, thus  generalizing previous work \cite{white} on
the subject. This was then generalized \cite{kuirouk1} for the case
of the generalized Grad-Shafranov equation (GGSE) (equation \eqref{eq1} below), i.e. using classical
Lie-group methods to find a one-parameter group admitted by the equation
and the corresponding group-invariant solution.

The Solov\'ev solution \cite{solov} and the Herrnegger-Maschke solution \cite{herne}
are among the most widely employed analytical solutions of the GSE, the
former corresponding to toroidal current density non-vanishing on the
plasma boundary while the later to toroidal current density vanishing
thereon. In the presence of flow, which plays a role in the transition
to improved confinement modes in tokamaks, the equilibrium satisfies
a GGSE, in general coupled with a Bernoulli equation involving the
pressure \cite{moroz,tasso}. For incompressible flow, the density
becomes a surface quantity and we obtain the GGSE \cite{tasso,simin}.
Generalized Solov\'ev solutions of the GGSE were derived in \cite{simin,kaltsas1}  and asymptotic expansion solutions to the general linearized GGSE in \cite{atanasiu,kaltsas3}. Also, some analytical solutions were recently constructed for the axisymmetric Hall magnetohydrodynamics GSE with incompressible flows \cite{Giannis2024}.

A symmetry group of a partial differential equation (PDE) is a set of transformations of the independent or/and dependent variables that allows to generate new classes of solutions from a single solution known.
According to the classical Lie group method, a similarity reduction of a PDE is  to determine a one parameter group admitted by the equation. Then one seeks the group-invariant solutions called  self-similar solutions (cf.  \cite{olver} for a recent description of the method). 
 Symmetry properties of a system
of Euler-type equations were studied in \cite{cicogna1}. Symmetry properties and
solutions to the GSE and the GGSE were studied in \cite{cicogna2,kuirouk2},
whereas symmetry methods for differential equations are exposed in \cite{olver}.
Translationally symmetric force-free states are constructed in \cite{tassi}.
The literature on exact solutions to the GSE is extensive and a brief reference
list includes \cite{carthy,cerfon}, while other symmetry and similarity reduction methods were employed in \cite{pou,frewer,kaltsas2,lukin}.

Aim of the present  study is to construct new families of analytic solutions to the GGSE by using an alternative  method of similarity reduction along the lines of \cite{litvi}, describing equilibrium configurations relevant to experimental fusion and space plasmas. Accordingly, a PDE can be reduced to an ordinary differential  equation  by employing an appropriate functional relation involving the dependent and independent variables, as  Eq. (4) below, after imposing certain conditions.

 The structure of the paper is as follows: In section 2 we present the direct method for finding the similarity reduction of the GGSE and show that the generalized Solov\'ev solution is a particular case of the above mentioned set of solutions. In section 3
we extend the set  of solutions to the GGSE and show that all group-invariant
solutions of \cite{kuirouk1} are in fact particular cases of the  extended set of solutions.
In section 5 we present a new similarity reduction of the GGSE,
not considered in \cite{litvi}, associated with solutions which can describe part of planetary magnetospheres.  
Finally, in section 5 we present a brief discussion on the novel results of this paper.


\section{Axisymmetric  equilibria with non-parallel flow from similarity reduction}\

We consider the GGSE in its
completely dimensionless form \cite{tasso,simin}
\bea
\label{eq1}
u_{rr}-\frac{1}{r}u_{r}+u_{zz}+\frac{1}{2}\frac{d}{du}
\left[\frac{X^{2}}{1-M_{p}^{2}}\right]+r^{2}\frac{dP_{s}}{du}
+\frac{1}{2}r^{4}\frac{d}{du}
\left[\rho\left(\frac{d\Phi}{du}\right)^{2}\right]=0\,.
\eea
Here, employing cylindrical coordinates($z,r,\phi$),  the function $u(r,z)$ relates to the poloidal magnetic flux function and labels the magnetic surfaces; the flux function $X(u)$ relates to the toroidal magnetic field (cf. Eq. (\ref{Xi}) below); $P_s(u)$ is the pressure in the absence of flow; $\rho(u)$ is the density; $M_p(u)$ is the Mach function of the poloidal velocity with respect to the respective Alfv\'en velocity; and $\Phi(u)$ is the electrostatic potential related to  the electric field. The flux functions $X(u)$, $M_p(u)$, $P_s(u)$, $\rho(u)$ and $\Phi(u)$ remain arbitrary. 
Also, the magnetic field and the velocity are given by the relations
\beq
\label{Bv}
\bB=I\nabla\phi+\left(1-M_p^2\right)^{-1/2}\nabla \phi \times \nabla u, \ \ 
\bv=\frac{M_p}{\sqrt{\rho}}\bB - r^2\left(1-M_p\right)^{1/2}\Phi^\prime \nabla \phi
\ee
where
\beq
\label{Xi}
I=\frac{X}{\left(1-M_p\right)}-r^2\frac{\sqrt{\rho}M_p\Phi^\prime}{\left(1-M_p\right)^{1/2}}
\ee

Compared with the GSE  the GGSE has the additional $R^4$-term related to the non parallel to the magnetic field component of the velocity through the electrostatic potential $\Phi$. This term can significantly modify the equilibrium; for example, in the presence of this term, the well known Solov\'ev equilibrium \cite{solov},  which is similar to the configuration of Fig. 1,  acquires an addition X-point outside the separatrix  (cf. Fig. 11 of \cite{simin}).   Also, although plasma rotation in magnetic confinement systems is low, i.e. $M_p\sim 10^{-2}$, the respective electric field and its shear play an important role in the transitions to improved confinement regimes, as the L-H-transition \cite{bur}.

\begin{figure}[!h]
\begin{center}
\includegraphics[scale=0.5]{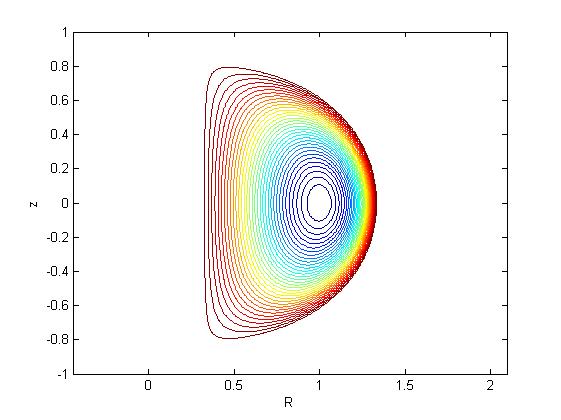}
\caption{The generalized Solov\'ev equilibrium in connection with the solution \eqref{eq3} of the GGSE for $E=-2$,
$G=0.2$, $F=-4$. The bounding flux surface corresponds to
$U_{b}=0.34$.}
 \label{fig1}
 \end{center}
 \end{figure}

Choosing the flux function terms 
to be constant, i.e.,  ($(1/2)d/du[X^2(1-M_p^2)]=G$, $dP_s/du=F$, $(1/2)d/du[\rho(d\Phi/du)^2]=E$ with $G$, $F$, $E$ constants)  we have the following form of the 
GGSE that will be the focus of this study
\bea
\label{eq2}
u_{rr}-\frac{1}{r}u_{r}+u_{zz}+Er^{4}+Fr^{2}+G=0\,.
\eea
Equation \eqref{eq2} is satisfied by  the generalized Solov\'ev-type solution \cite{simin}
\bea
\label{eq3}
u_{Sol}=
\left[z^{2}\left(r^{2}-\frac{G}{2}\right)-\frac{(F+E+2)}{8}(r^{2}-1)^{2}-\frac{E}{24}(r^{2}-1)^{3}\right]\,.
\eea
The respective equilibrium configuration  inside the separatrix, up-down symmetric with a pair of X-points,  is plotted in Fig. \ref{fig1}. For $E=0$,  (\ref{eq3}) reduces to the well known Solov\'ev solution \cite{solov}.

\begin{figure}[!h]
\begin{center}
\includegraphics[scale=0.5]{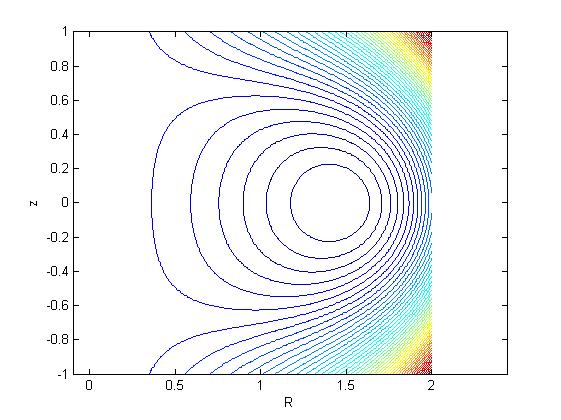}
\caption{The equilibrium configuration determined by  \eqref{eq4} in conjunction with \eqref{eq9} and  \eqref{eq10}.}
 \label{fig2}
 \end{center}
 \end{figure}

We consider now the similarity reduction of Eq. \eqref{eq2} via
the ansatz \cite{litvi}
\bea
\label{eq4}
u(r,z)=\alpha(r,z)+\beta(r,z)w[x(r,z)]\,.
\eea
Substituting  \eqref{eq4} into  \eqref{eq2} we obtain
\bea
\label{eq5}
\beta\left [(x_{r})^{2}+(x_{z})^{2}\right]w^{''}&+&
\left[\left(2\beta_{r}-\frac{1}{r}\beta\right)x_{r}+2\beta_{z}x_{z}+\beta(x_{rr}+x_{zz})\right]w^{'}
+\nn \\&+&\left(\beta_{rr}+\beta_{zz}-\frac{1}{r}\beta_{r}\right)w+\nn
\\&+&(\alpha_{rr}+\alpha_{zz}-\frac{1}{r}\alpha_{r}+Er^{4}+Fr^{2}+G)=0\,.
\eea
Let us choose as a first illustration of the method the case
where $\alpha=\alpha(r)$, $\beta=\beta(r)$ and $x=x(z)$. 
Then Eq. \eqref{eq5} becomes
\bea
\label{eq6}
\beta x_{zz}+(\beta_{rr}-\frac{1}{r}\beta_{r})x+
(\alpha_{rr}-\frac{1}{r}\alpha_{r}+Er^{4}+Fr^{2}+G)=0\,.
\eea
This equation can be satisfied by choosing $x(z)=z^{2}/2$, provided that 
the functions $\alpha , \beta$ satisfy the ODEs
\bea
\label{eq7}
\beta_{rr}-\frac{1}{r}\beta_{r}=0\,,
\eea
\bea
\label{eq8}
\beta+\alpha_{rr}-\frac{1}{r}\alpha_{r}+Er^{4}+Fr^{2}+G=0\,.
\eea
These equations are solved for
\bea
\label{eq9}
\beta(r)=\beta_{0}+\beta_{2}r^{2}\,,
\eea
and
\bea
\label{eq10}
\alpha(r)=\alpha_{0}+[\alpha_{2}+\frac{1}{4}(\beta_{0}+G)]r^{2}
-\frac{1}{8}(\beta_{2}+F)r^{4}-\nn \\-\frac{E}{24}r^{6}-
\frac{1}{2}(\beta_{0}+G)r^{2}\ln r\,.
\eea
The solution $u(r,z)$ of the GGSE \eqref{eq2} by means of \eqref{eq4}, \eqref{eq9} and \eqref{eq10} is plotted in Fig. \ref{fig2}, for $E=-2$, $G=0.2,$ $F=-4$,
$\alpha_{0}=1$, $\alpha_{2}=-1$, $\beta_{0}=-0.1$, and $\beta_{2}=4.1$.
Compared with the generalized Solov\'ev solution of Fig. \ref{fig1} the equilibrium has an external part of open magnetic surfaces and the closed magnetic surfaces in the inner  part have less elongation. 
The generalized Solov\'ev
solution  \eqref{eq3} is obtained from  \eqref{eq4}
by choosing 
$\beta_0=-G$, $\beta_2=2$, $a_{0}=-(F+2E/3+2)/8$ and $a_{2}=(F+E/2+2)/4$.

Returning to Eq. \eqref{eq5},  setting $x=z/r$ and assuming
$\alpha=\alpha(r,z)$ and $\beta=\beta(r)$ we obtain
\bea
\label{eq12}
\beta(1+x^{2})w^{''}&+&(3\beta-2r\beta_{r})xw^{'}+
(r^{2}\beta_{rr}-r\beta_{r})w+\nn \\&+&
r^{2}(\alpha_{rr}+\alpha_{zz}-\frac{1}{r}\alpha_{r}+Er^{4}+Fr^{2}+G)=0\,.
\eea
This equation is satisfied by $\beta(r)=r^{\nu}$ and by any solution
of the following equations
\bea
\label{eq13}
(1+x^{2})w^{''}(x)+(3-2\nu)xw^{'}(x)+\nu(\nu-2)w(x)=0\,,
\eea
\bea
\label{eq14}
\alpha_{rr}+\alpha_{zz}-\frac{1}{r}\alpha_{r}+Er^{4}+Fr^{2}+G=0\,.
\eea
Equations \eqref{eq13} and \eqref{eq14} constitute a new similarity
reduction of the GGSE. Here we note an error in Eq. (15) of \cite{litvi}
which in {\it not} a solution of Eq. (13) of \cite{litvi}
neither for $\nu=1/2$ nor for $\nu=3/2$.

 Eq. \eqref{eq13} has the analytic solution \cite{litvi}
\bea
\label{eq18}
w(x)&=&c_{1}(1+x^{2})^{\frac{(2\nu -1)}{4}}P_{1/2}^{\frac{(1-2\nu)}{2}}(ix)+
\nn \\&+&c_{2}(1+x^{2})^{\frac{(2\nu -1)}{4}}Q_{1/2}^{\frac{(1-2\nu)}{2}}(ix)\,,
\eea
in terms of the associated Legendre functions $P_{1/2}^{(1-2\nu)/2}(ix)$ and $Q_{1/2}^{(1-2\nu)/2}(ix)$. These functions  involve infinite converging  series; only for particular values of $\nu$, e.g. $\nu=0,1,2$ to be considered  below, the solution  is expressed in closed form, i.e. in terms of Legendre polynomials. For this reason, in the former case,  instead of employing the analytic solution we preferred to solve Eq. (\ref{eq13}) numerically.

\begin{figure}[!h]
\begin{center}
\includegraphics[scale=0.5]{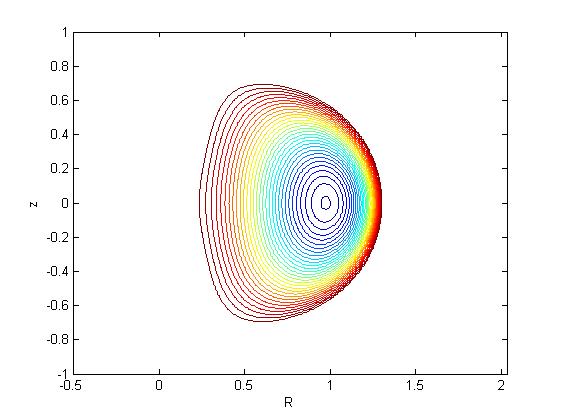}
\caption{The up-down symmetric equilibrium configuration  with smooth bounding surface in connection to the solution (\ref{eq16}) of the GGSE  (\ref{eq2}).}
 \label{fig3}
 \end{center}
 \end{figure}

In particular, for $\nu=1$ we obtain the solution of Eq. (ref{eq13}) in the
following two equivalent forms 
\bea
\label{eq15}
w&=&c_{1}[x+\sqrt{x^{2}+1}]+c_{2}[x+\sqrt{x^{2}+1}]^{-1}\,,\nn \\
w&=&d_{1}\sqrt{x^{2}+1}+d_{2}x\,,
\eea
where $c_{1}, c_{2}, d_{1}, d_{2}$ are arbitrary integration
constants. The solution for $\nu=1$ is given by
\bea
\label{eq16}
u(r,z)=u_g(r,z)+r[c_{1}[x+\sqrt{x^{2}+1}]+c_{2}[x+\sqrt{x^{2}+1}]^{-1}]\,,
\eea
where $u_g(r,z)$ is any solution of the GGSE such as the
generalized Solov\'ev solution (\ref{eq3}) and $x=z/r$. It is plotted in Fig. \ref{fig3},
for $E=-2,$ $G=0.2$, $F=-4,$ $c_{1}=c_{2}=0.05$. The bounding flux
surface corresponds to $U_{b}=0.40$. 
Although the configuration is similar to that of Fig. \ref{fig1}, it does not have X-points.

\begin{figure}[!h]
\begin{center}
\includegraphics[scale=0.5]{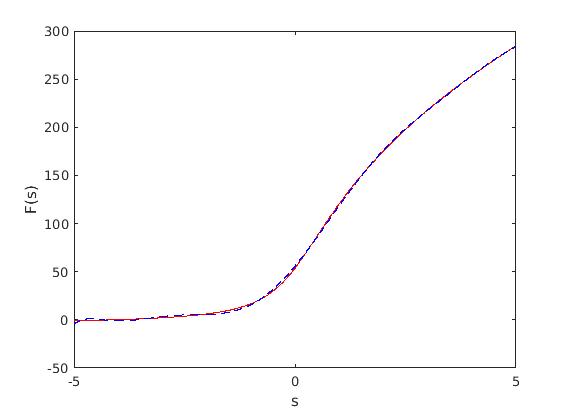}
\caption{The numerical solution of Eq. \eqref{eq13} for
$\nu=1/2$ (red solid line) versus the polynomial fit (dashed blue line).}
 \label{fig4}
 \end{center}
 \end{figure}

For $\nu=1/2$ and $\nu=3/2$ the solution of Eq. (\ref{eq13}) is expressed in terms of infinite series; thus, as we already noted earlier, we preferred solving Eq. \eqref{eq13} numerically. The solutions   are plotted in Figs. 4, 5, 6 and 7. Specifically,  for $\nu=1/2$ we integrated numerically Eq. \eqref{eq13}
in the interval $-5\leq x\leq 5$ for $w(-5)=-1$ and $w^{'}(-5)=1$. Numerical
fitting with a polynomial of ninth order gives
\bea
w(x)=0.0002x^9-0.0008 x^8-0.0151x^7+0.0481x^6
+0.3697x^5\nn\\
-1.0552 x^4-4.4810 x^3+12.2627 x^2+55.5659 x+55.8746\,.\nn
\eea
This is plotted in Fig. \ref{fig4}.
The respective solution of Eq. \eqref{eq4} for
$E=-2$, $G=0.2$, $F=-4$, is plotted in Fig. 5, where $u=u_{Sol}+cr^{1/2}w$ with $u_{Sol}$ being the generalized Solov\'ev solution \eqref{eq3} and 
$c=5\times 10^{-4}$.  The configuration  is D-shaped up-down asymmetric. 
\begin{figure}[!h]
\begin{center}
\includegraphics[scale=0.5]{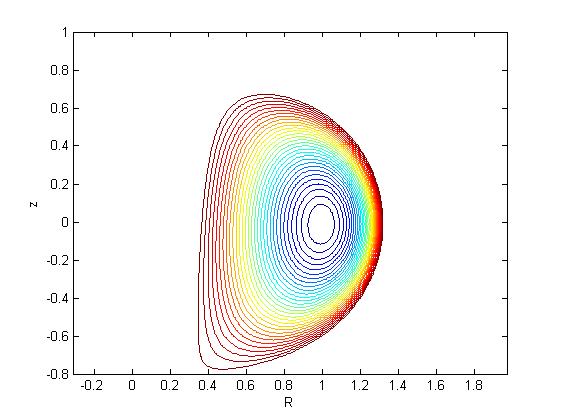}
\caption{Up-down asymmetric equilibrium determined by  \eqref{eq4} with the aid of the numerical solution of Eq. \eqref{eq13} for $w(x)$ with $\nu=1/2$, shown in  Fig. 4. }
 \label{fig5}
 \end{center}
 \end{figure}
The same procedure applies for the case $\nu=3/2$.
We integrated numerically Eq. \eqref{eq13} in the interval $-5\leq x\leq 5$
for $w(-5)=-1$ and $w^{'}(-5)=1$. Numerical fitting with a quintic polynomial
gives
\bea
w(x)=-0.0037x^5-0.0178x^{4}+0.2269x^{3}+1.8353x^{2}+5.2115x+6.8747\,.\nn
\eea
This is plotted in Fig. \ref{fig6}.
\begin{figure}[!h]
\begin{center}
\includegraphics[scale=0.5]{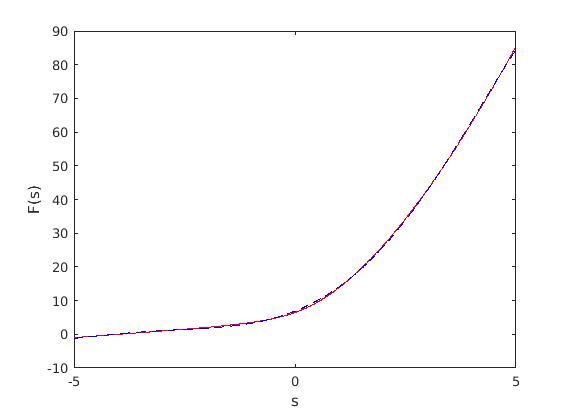}
\caption{The numerical solution of Eq. \eqref{eq13} for
$\nu=3/2$ (red solid line) versus the polynomial fit (dashed blue line).}
\label{fig6}
 \end{center}
 \end{figure}
 The solution of Eq. \eqref{eq4} for
$E=-2$, $G=0.2$, $F=-4$, is plotted in Fig. \ref{fig7}, where $u=u_{Sol}+cr^{3/2}w$
with $c=5\times 10^{-3}$. 
\begin{figure}[!h]
\begin{center}
\includegraphics[scale=0.5]{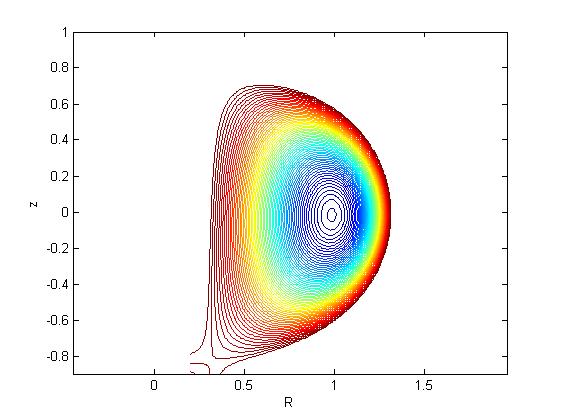}
\caption{Up-down asymmetric equilibrium with a lower X-point 
 determined by  \eqref{eq4} with the aid of the numerical solution of Eq. \eqref{eq13} for $w(x)$ with $\nu=3/2$, shown in  Fig. \ref{fig6}. }
 \label{fig7}
 \end{center}
 \end{figure}
This is a diverted equilibrium with a lower X-point.   

For $\nu=3$ we obtain the explicit solution of Eq. \eqref{eq13} as
\bea
\label{eq17}
w=c_{1}[(x^{2}-2)\sqrt{x^{2}+1}+3x\sinh^{-1}(x)]+c_{2}x\,.
\eea

\section{More equilibria from similarity reduction}\

New non-trivial solutions to the GGSE can be generated from any
solution to Eq. \eqref{eq14}. For example,  we will employ the following solutions to Eq. \eqref{eq14}: 
\bea
\label{eq19}
\alpha(r,z)=-\frac{E}{24}r^{6}-\frac{1}{8}Fr^{4}-\frac{1}{2}Gz^{2}\,,
\eea
\bea
\label{eq20}
\alpha(r,z)=-\frac{E}{24}r^{6}-\frac{1}{8}Fr^{4}+\frac{G}{2}r^{2}\ln r-Gz^{2}\,.
\eea
Using Eq.  \eqref{eq18} for $\nu=3$ and \eqref{eq19}  we have the solution
\bea
\label{eq21}
u(r,z)&=&c_{1}[(z^{2}-2r^{2})\sqrt{r^{2}+z^{2}}+3r^{2}z\sinh^{-1}(\frac{z}{r})]+\nn \\
&+&c_{2}r^{2}z-\frac{E}{24}r^{6}-\frac{1}{8}Fr^{4}-\frac{1}{2}Gz^{2}\,.
\eea
Yet another way to produce new solutions is to
consider $\alpha=\alpha(r)$ in Eq. \eqref{eq14} leading to the solution
\bea
\label{eq22}
\alpha(r)=d_{1}r^{2}+d_{2}-\frac{E}{24}r^{6}-\frac{F}{8}r^{4}-\frac{1}{2}Gr^{2}\ln r\,.
\eea
In this case  using again the solution \eqref{eq18} for $\nu=3$ we find from \eqref{eq4}
\bea
\label{eq23}
u(r,z)=c_{1}[(z^{2}-2r^{2})\sqrt{r^{2}+z^{2}}+3r^{2}z\sinh^{-1}(\frac{z}{r})]+
c_{2}r^{2}z+\alpha_{1}(r)\,.
\eea

\begin{figure}[!h]
\begin{center}
\includegraphics[scale=0.5]{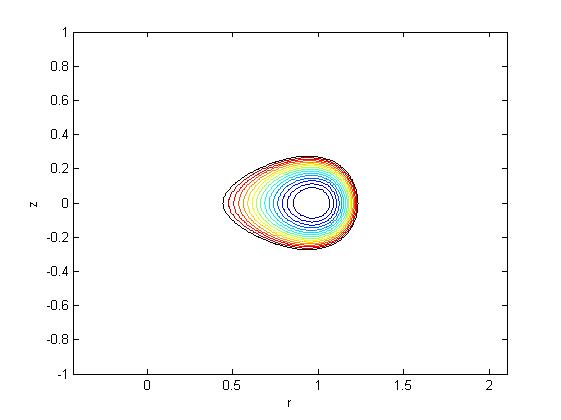}
\caption{Equilibrium with negative triangularity associated with the solution  \eqref{eq24}.
}
 \label{fig8}
 \end{center}
 \end{figure}

Moreover if $u_g(r,z)$ is a solution to the GGSE  \eqref{eq2} then
$u_g(r,z)+\beta(r,z)w[x(r,z)]$ is also a solution, thus making it possible
to combine solutions with different $\nu$. As an  example, combining
the solutions for $\nu=3$ with those for $\nu=1$ presented previously  and Eq. \eqref{eq19}
we find
\bea
\label{eq24}
u(r,z)&=&c_{1}\left[(z^{2}-2r^{2})\sqrt{r^{2}+z^{2}}+3r^{2}z\sinh^{-1}\left(\frac{z}{r}\right)\right]
+\nn \\&+&c_{2}r^{2}z-\frac{E}{24}r^{6}-\frac{F}{8}r^{4}-\frac{1}{2}Gz^{2}\nn
\\&+&c_{3}[z+\sqrt{r^{2}+z^{2}}]+c_{4}\frac{r^{2}}{z+\sqrt{r^{2}+z^{2}}}\,.
\eea
This is plotted in Fig. \ref{fig8}, for $E=-2$, $G=-2.2$, $F=-4$, $c_{1}=0.4$,
$c_{2}=c_{3}=c_{4}=0$. The bounding flux surface corresponds to $U_{b}=-0.05$
while at the center  we have $U_{c}=-0.2$. 
Note that this equilibrium configuration has inverse D-shape. Such equilibria has been the subject of recent tokamak studies because they may exhibit  improved confinement properties \cite{au,me}.

The family of solutions above contains the
classical and non-classical group-invariant solutions of the GGSE so
that it greatly extends the range of available analytical solutions. It is
interesting that all group-invariant solutions to the GGSE are in fact particular
cases of the above similarity reduction. Indeed, Eq. \eqref{eq13}
with $\nu=0$ is solved for
\bea
\label{eq25}
w(x)=\frac{c_{1}x}{\sqrt{1+x^{2}}}+c_{2}\,.
\eea
Adding a particular solution by Eq.  \eqref{eq22} we are led to
\bea
\label{eq26}
u_{3}(r,z)=c_{1}\frac{z}{\sqrt{r^{2}+z^{2}}}+c_{2}+\frac{1}{2}Gr^{2}(1-\ln r)
-\frac{1}{8}Fr^{4}-\frac{E}{24}r^{6}\,.
\eea
This is the $U^{(3)}_{inv}(x,y)-$solution of \cite{kuirouk1} given by Eq. (23) therein. \newline
For $\nu=1$ a solution of \eqref{eq13} is the second of \eqref{eq15}:
\bea
\label{eq27}
w(x)=c_{1}\sqrt{1+x^{2}}+c_{2}x\,,
\eea
while for $\nu=-1$ we solve Eq. \eqref{eq13} for
\bea
\label{eq28}
w(x)=\frac{c_{1}}{(1+x^{2})^{3/2}}+c_{2}
\left[\frac{x}{1+x^{2}}+\frac{\sinh^{-1}(x)}{(1+x^{2})^{3/2}}\right]\,.
\eea
Taking a particular linear combination of \eqref{eq27}) and \eqref{eq28}
and  $a(r)$ from Eq. \eqref{eq22} leads to
\bea
\label{eq29}
u_{1}(r,z)&=&c_{1}\sqrt{r^{2}+z^{2}}+c_{2}\frac{r^{2}}{(r^{2}+z^{2})^{3/2}}+\nn
\\&+&\frac{1}{3}Gr^{2}-\frac{E}{24}r^{6}-\frac{F}{8}r^{4}-\frac{1}{2}Gr^{2}\ln r\,.
\eea
This is the $U^{(1)}_{inv}(x,y)-$solution of \cite{kuirouk1} given by Eq. (21) therein. \newline
Finally, Eq. \eqref{eq13} with $\nu=2$ is solved for
\bea
\label{eq30}
w(x)=c_{1}x\sqrt{1+x^{2}}+c_{1}\sinh^{-1}(x)+c_{2}\,.
\eea
Using $\alpha(r,z)$ from Eq. \eqref{eq20} we have
\bea
\label{eq31}
u_{2}(r,z)&=&c_{1}z\sqrt{r^{2}+z^{2}}+c_{1}r^{2}\sinh^{-1}\left(\frac{z}{r}\right)
+c_{2}r^{2}-\nn \\&-&Gz^{2}-\frac{E}{24}r^{6}-\frac{F}{8}r^{4}+\frac{G}{2}\ln r\,,
\eea
which is the $U^{(2)}_{inv}(x,y)-$solution of \cite{kuirouk1} given by Eq. (22) therein.

\section{Equilibria from a new similarity reduction}\

\begin{figure}
\begin{center}
\includegraphics[scale=0.5]{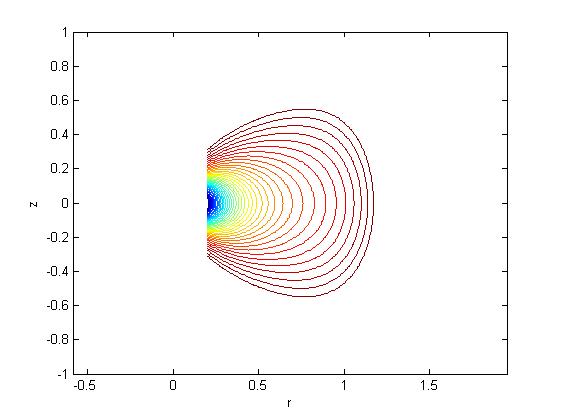}
\caption{Equilibrium configuration in connection  with the solution  \eqref{eq35} pertinent to part of  a planetary magnetosphere. }
 \label{fig9}
 \end{center}
 \end{figure}

We consider now a new similarity reduction not considered in \cite{litvi}.
Returning to Eq. \eqref{eq13} and setting $x=(r^{2}+z^{2})/r$,
$\alpha=\alpha(r,z)$, $\beta=\beta(r)$ with $2\beta_{r}-(1/r)\beta=0$, 
 implying that 
 $\beta=\beta_{0}r^{1/2}$, we obtain
\bea
\label{eq32}
x^{2}w^{''}+2xw^{'}-\frac{3}{4}w=0\,,
\eea
\bea
\label{eq33}
\alpha_{rr}+\alpha_{zz}-\frac{1}{r}\alpha_{r}+Er^{4}+Fr^{2}+G=0\,,
\eea
the latter equation being identical with \eqref{eq14}.
Eq. \eqref{eq32} is solved for
\bea
\label{eq34}
w(x)=c_{1}\sqrt{x}+\frac{c_{2}}{x^{3/2}}\,.
\eea
Using the solution \eqref{eq20} of Eq. \eqref{eq14} for $\alpha(r)$ we find  the solution
\bea
\label{eq35}
u(r,z)&=&-\frac{E}{24}r^{6}-\frac{F}{8}r^{4}+\frac{G}{2}r^{2}\ln r-Gz^{2}+
\nn \\&+&r^{1/2}
\left[c_{1}\left(\frac{r^{2}+z^{2}}{r}\right)^{1/2}+c_{2}\left(\frac{r^{2}+z^{2}}{r}\right)^{-3/2}\right]\,.
\eea
This is plotted in Fig. \ref{fig9}, for $E=-2,$ $G=-1.2$, $F=-4$, $c_{1}=c_{2}=-1$.
Such a solution could be useful in modelling parts of planetary magnetospheres, e.g., the earth's magnetosphere, for distances not far from the planet surface \cite{bo,pou,lukin} and also emerging flux ropes in the solar corona.



%

%


\section{Conclusions}

We have considered a similarity reduction of the GGSE along the lines presented
in \cite{litvi}. New classes of exact solutions to the GGSE were produced, with
the generalized Solov\'ev solution being a particular case, describing a variety of D-shaped equilibrium configurations  exhibiting up-down symmetry or asymmetry and either positive or negative triangularity and diverted configurations with a lower X-point. All group-invariant
solutions of the GGSE presented in \cite{kuirouk1} are shown to be particular
cases of the above formalism. In addition, a new class of solutions, not considered in
\cite{litvi}, has been constructed  which can describe part of a planetary magnetoshphere for short distances from the planet surface.


\section*{Aknowledgments}

The authors would like to thank the referees for constructive comments to improve the paper.
This work has received funding from the National Fusion
Programme of the Hellenic Republic$-$General Secretariat for
Research and Innovation.


\end{document}